\documentclass[structabstract]{aa}  
\usepackage{graphicx}
\usepackage{natbib}
\bibpunct{(}{)}{;}{a}{}{,} 
\usepackage{txfonts}
\begin{document}
   \title{Photospheric flux cancellation and associated flux rope formation and
          eruption}

   \author{L. M. Green\inst{1}
     \and  B. Kliem\inst{1,2,3}
     \and  A. J. Wallace\inst{1}
          }

   \institute{University College London, Mullard Space Science Laboratory,
              Holmbury St.\ Mary, Dorking, Surrey, RH5 6NT, UK 
         \and University of Potsdam, Institute of Physics and Astronomy,
              14476 Potsdam, Germany
         \and Naval Research Laboratory, Space Science Division,
              Washington, DC 20375, USA
             }

   \date{Received 3 June 2010 / Accepted 18 October 2010}

 
  \abstract
   {}
   {We study an evolving bipolar active region that exhibits flux
    cancellation at the internal polarity inversion line, the
    formation of a soft X-ray sigmoid along the inversion line and a
    coronal mass ejection. The aim is to investigate the quantity of
    flux cancellation that is involved in flux rope formation in the
    time period leading up to the eruption.}
   {The active region is studied using its extreme ultraviolet and
    soft  X-ray emissions as it evolves from a sheared arcade to flux
    rope configuration. The evolution of the photospheric magnetic
    field is described and used to estimate how much flux is 
    reconnected into the flux rope.}
   {About one third of the active region flux cancels at the internal
    polarity inversion line in the 2.5~days leading up to the
    eruption. In this period, the coronal structure evolves from a
    weakly to a highly sheared arcade and then to a sigmoid that
    crosses the inversion line in the inverse direction. These
    properties suggest that a flux rope has formed prior to the eruption.
    The amount of cancellation implies that up to 60\% of the active
    region flux could be in the body of the flux rope. We point out
    that only part of the cancellation contributes to the flux in the
    rope if the arcade is only weakly sheared, as in the first part of
    the evolution. This reduces the estimated flux in the rope to
    $\sim\!30\%$ or less of the active region flux. We suggest that
    the remaining discrepancy between our estimate and the limiting
    value of $\sim\!10\%$ of the active region flux, obtained
    previously by the flux rope insertion method, results from the
    incomplete coherence of the flux rope, due to nonuniform
    cancellation along the polarity inversion line.
    A hot linear feature is observed in the active region which
    rises as part of the eruption and then likely traces out field
    lines close to the axis of the flux rope. The flux cancellation
    and changing magnetic connections at one end of this feature suggest
    that the flux rope reaches coherence by reconnection shortly before
    and early in the impulsive phase of the associated flare.
    The sigmoid is destroyed in the eruption but reforms quickly,
    with the amount of cancellation involved being much smaller than in
    the course of its original formation.}
   {}


   \maketitle
%

\section{Introduction}

Photospheric magnetic flux cancellation is observed as the collision 
and subsequent disappearance of small scale opposite polarity 
magnetic fragments in magnetograph data \citep{martin85}. It is
associated with the submergence of small scale loops, which have a
small radius of curvature and can be pulled under the photosphere by 
the magnetic tension force \citep{harvey99}. These small loops are
likely to be the product of magnetic reconnection which sets in as the
opposite polarity fragments collide \citep{zwaan87, vanballegooijen89}. 
Recent work has
supported the location of the reconnection associated to flux
cancellation to be in the photosphere \citep{Yurchyshyn&Wang2001,
BellotRubio&Beck2005}, but reconnection at higher levels may also be
relevant \citep{Wang&Muglach2007}. Above the small, submerging loop,
magnetic field with a concave-up shape is formed. Depending on the
plasma beta in the locality and on the strength of the overlying
field, the concave-up field lines will be line-tied and form a
so-called bald patch \citep{titov&al93}, or they can relax upward 
to form a coronal loop.

Cancelling magnetic features are observed along photospheric polarity
inversion lines (PIL) throughout the quiet sun, at the periphery of active
regions and in active regions. Sustained reconnection and flux
cancellation along PILs is of particular interest as it is a mechanism
by which helical field lines can be formed from a sheared arcade,
building up a flux rope \citep{vanballegooijen89}. 

The study of magnetic flux ropes is relevant to many areas of solar
physics and since we are currently unable to directly observe the
coronal  magnetic field, other observational support for the
occurrence of flux ropes must be found. Such support is given if the
horizontal component of vector magnetic field data is inversely directed
at a bald patch section of the PIL. The concave-up
configuration could be produced by field lines at the bottom of a
flux rope \citep{athay83,lites05,canou09}, particularly if the region
is bipolar, so that a double arcade can be excluded. Support for a flux
rope is also given if continuous S shaped (sigmoidal) sources of X-ray
and EUV emission, which follow the magnetic field lines, exhibit an
inverse crossing of the PIL. This, however, is conclusive only if
the sigmoid survives an eruption \citep{gibson06, green09}, since S
shaped field lines in a sheared arcade \citep{Antiochos&al1994} can only
then be excluded.

Numerical simulations of the bodily emergence of a flux rope indicate
that the process essentially stops as the magnetic axis of the rope
reaches the photosphere, producing a sheared arcade in the corona 
\citep{fan01}. In the situation where the flux rope axis is unable to
cross the photosphere, flux cancellation may be an important 
mechanism by which a stable flux rope can form in situ in the solar
atmosphere \citep{Amari&al2003b, Amari&al2010, aulanier10}.
Observations of flux cancellation
can therefore allow the investigation of the location and timing of
flux rope formation, and can help answer fundamental questions related
to the energy accumulation timescale and the flux content of the rope.
These are important questions related to the general evolution of the
sun's magnetic field and to the onset and driver of eruptive events. 

Sigmoidal regions are sites of flux rope formation \citep{green09,
Tripathi&al2009} and have a high likelihood of producing an eruption
\citep{canfield99} and so observations of flux cancellation in these
regions can be used to investigate both the flux rope formation and
the evolution toward a loss of equilibrium resulting in a coronal mass
ejection (CME). The loss of equilibrium can be described as an ideal
MHD instability \citep{hood81,kliem06}, as a catastrophe 
\citep{forbes91}, or as a force imbalance between flux rope and
overlying arcade field \citep{mackay06}. The overlying arcade field
provides a downward tension force, whilst the Lorentz force in the
flux rope points upward. The arcade tension force reduces, whilst the
flux rope force grows as the sheared arcade field transforms into a
flux rope. A force-free equilibrium of the flux rope/arcade
configuration is therefore only possible up to a limiting value of the
ratio between these fluxes. This value depends on the configuration of
photospheric flux and coronal currents. For four decayed active regions,
one of which contains a sigmoid, \cite{bobra08}, \cite{Su&al2009} 
and \cite{savcheva09} found that
the force balance will be lost if the axial flux in the rope surpasses a
threshold lying in the range $\approx\!(10\mbox{--}14)\%$ of the active
region flux. Recent numerical simulations of eruptions driven by flux
cancellation have obtained similarly small values for the fraction of
cancelled flux at the onset of eruption, lying in the range
$(6\mbox{--}10)\%$ \citep{aulanier10, Amari&al2010}.
A related result is that, generally, a coronal flux rope is
unstable if the overlying field falls off sufficiently rapidly with
height \citep{vantend78, torok05, kliem06}.

Previously, dimming signatures of eruptions have been used to
investigate the flux content of the erupting flux rope and make a 
link with in situ observations of flux ropes at 1AU 
\cite[e.g.][]{webb00, mandrini05, attrill06, jian06}. However, the
erupting stucture is strongly modified by magnetic reconnection that
occurs in the current sheet under the rising flux rope and produces 
the post-eruption flare arcade \cite[e.g.,][]{qiu07}. We rather use 
flux cancellation in the active region to study the flux content of
the structure before the eruption onset.

In a small number of cases, the eruption from a sigmoidal active
region has been observed as the rise of a faint, nearly linear feature in the
soft X-ray emission, suggested to be formed near the magnetic axis of
a flux rope \citep{moore01,mckenzie08}. Alternatively, a recent
simulation of an eruption from a sigmoidal active region suggests that
this feature may be formed in the current layer that develops
in the arcade field above an erupting flux rope \citep{aulanier10}. 

Here we present observations of a sigmoidal active region
where the main phases of the evolution of the magnetic field are
observed; from emergence to decay. We investigate the formation 
of a flux rope via reconnection and cancellation
along the internal PIL, and the point at which the overlying
arcade field is unable to hold down the flux rope leading to a loss of
equilibrium and coronal mass ejection. The evolution of a linear 
feature observed before and during the eruption is also studied.


\section{Photospheric field evolution}
\label{sect:mdi}

   \begin{figure}
   \centering
   \resizebox{\hsize}{!}{\includegraphics{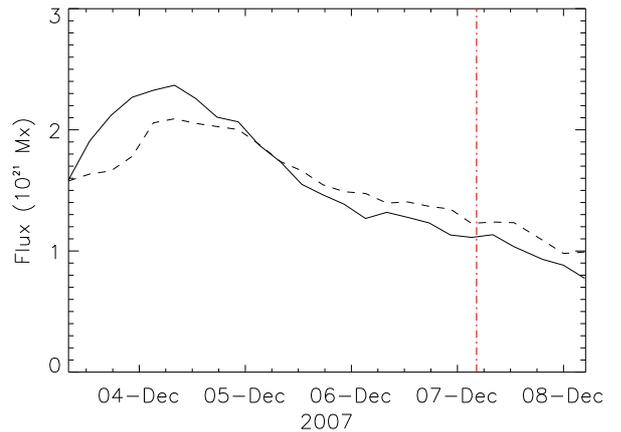}}
      \caption{Flux evolution in AR~10977 as determined by the
               \textsl{SOHO}/MDI data. The positive (negative) flux is
               shown by the continuous (dashed) line and the time of
               the eruption by the red dash-dot line.}
         \label{fig:mdigraph}
   \end{figure}

\textsl{SOHO}/MDI measures the line of sight magnetic field in the
mid-photosphere \citep{scherrer95} and shows that NOAA active region
(AR) 10977 emerges in a quiet sun region which has a dominantly
negative polarity. 

The evolution of the photospheric flux is monitored by fitting a
separate contour around each polarity which is adjusted with time to
account for the spatial evolution of the active region. The contour is
defined by eye and the data are displayed in such as way as to ensure
weak flux fragments can be determined. The data are corrected for the
area foreshortening that occurs away from central meridian and the
radial field component is estimated using the IDL Solar Software
routine \texttt{zradialise}. The latter part of the emergence phase is
observed and  lasts until around 4~December 06:00~UT. During this time
the polarities are imbalanced with more positive than negative flux
detected. This is due to a projection effect produced by the presence
of a significant horizontal field component as the flux crosses the
photosphere which in the eastern (western) hemisphere results in the
following (leading) polarity of the active region dominating the flux
budget \citep{green03}. During the subsequent decay phase the photospheric field
evolution is dominated by fragmentation, motion due to supergranular
flows and cancellation of flux where opposite polarity fragments
collide. This flux cancellation involves field at the internal PIL of
the bipolar region as well as surrounding quiet sun field. The
dominance of negative polarity in the surrounding field results in a
stronger reduction of the region's positive flux than the negative as
the cancellation proceeds.

As the active region disperses the flux determined by using a
contour increasingly includes a contribution from small-scale fields
that  presumably connect within the boundary defined for each 
polarity, rather than across the PIL. This small-scale flux is
quantified by measuring the magntitude of positive (negative) flux
within the negative (positive) polarity contour.
Both polarities of the small-scale flux are then removed from the
budget for each contour. The small-scale flux contributes at
most 5\% of the active region flux. Approximately 50\% of the active
region flux cancels in the time period from the peak flux value on
4~December to the time of the eruption on 7~December
(Fig.~\ref{fig:mdigraph}). The dispersal of the region and
cancellation of its flux continue through the subsequent days.

   \begin{figure}
   \centering
   \resizebox{.9\hsize}{!}{\includegraphics{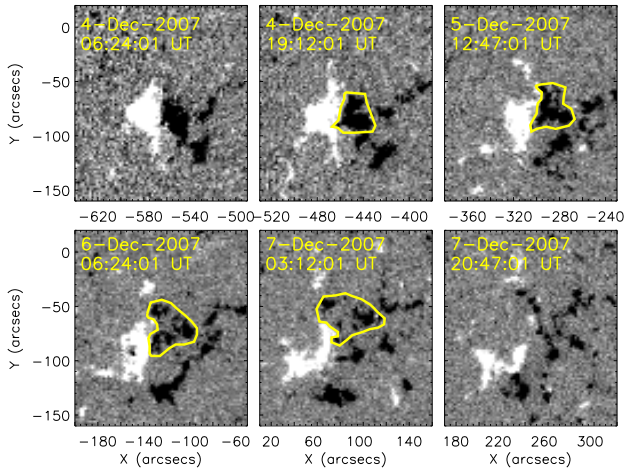}}
      \caption{Photospheric flux evolution in AR~10977 as determined
               with \textsl{SOHO}/MDI magnetograms. The data are
               displayed between $\pm\!100$~Gauss. The yellow contour
               shows the area within which the negative flux is
               measured during the main flux cancellation episode. In
               a similar way, the total unsigned flux of the region is
               determined by fitting a countour around the whole
               active region.}
         \label{fig:mdi}
   \end{figure}
 
\begin{table}
\caption{Flux evolution within the negative fragment that undergoes
  cancellation at the active region's internal polarity inversion
  line.}
\label{table:1}      
\centering                          
\begin{tabular}{c c}        
\hline\hline                 
Date & Flux (10$^{21}$ Mx) \\    
\hline                        
   4 Dec. 19:12 UT & -1.10  \\      
   5 Dec. 12:47 UT & -0.82 \\
   6 Dec. 06:24 UT & -0.58 \\
   7 Dec. 03:12 UT & -0.39 \\
\hline                                   
\end{tabular}
\end{table}

In order to investigate how much flux is involved in the building of
the flux rope, we consider now only the flux cancellation at the
internal PIL. The negative polarity is fragmented throughout the
active region's lifetime; however, one main fragment is observed to be
involved with ongoing cancellation at the internal PIL in the time
leading up to the eruption on 7~December (a fragment to the south
cancels thereafter). Figure~\ref{fig:mdi}, panels~2--5, shows this
fragment within the yellow contour. The cancellation occcurs
predominantly at the PIL under the sigmoid centre and can be tracked
reliably since the cancellation occurs with positive polarity
fragments of the active region and not with the surrounding, and 
dominantly negative, quiet sun field. The evolution of the active
region's positive flux cannot be followed in this way as it is not
possible to separate the cancellation with  surrounding flux from that
at the internal PIL. During the time period from 4~December 19:12 to
7~December 03:12~UT, $0.71\times10^{21}$~Mx of flux is cancelled
at the internal PIL
(Table~\ref{table:1}). This represents $\approx34\%$ of the peak
negative flux value of the active region. Since the cancellation is
enabled by reconnection, with flux remaining in the active region
(Sect.~\ref{s:discussion}), an amount of flux equal to that cancelled is
potentially available for transformation into the flux rope. 

At the time of the eruption, the unsigned average flux in the active
region was $1.17\times10^{21}$~Mx (see Fig.~\ref{fig:mdigraph}). This
flux value includes that in the flux rope body, which intersects the
photosphere at the flux rope feet, as well as that of the overlying
arcade field.


\section{Formation of sigmoid}
\label{sect:sigmoidformation}

  \begin{figure}
   \centering
   \resizebox{.9\hsize}{!}{\includegraphics{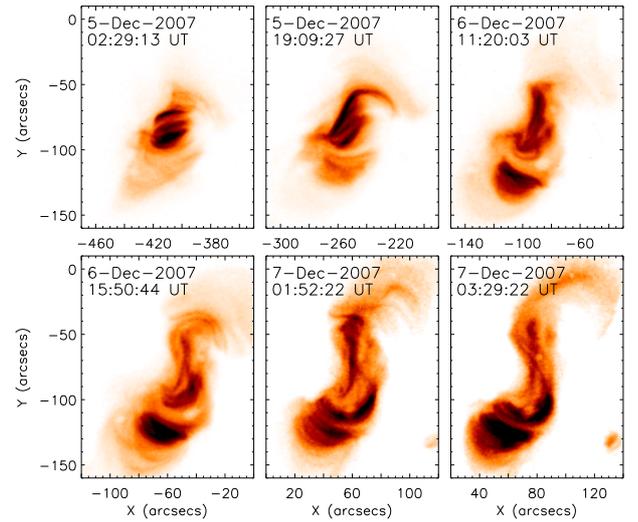}}
      \caption{\textsl{Hinode}/XRT C Poly filter observations of AR~10977 
               showing the formation of the sigmoid from an
               increasingly sheared arcade.}
         \label{fig:sigmoidform}
   \end{figure}

  \begin{figure}
   \centering
   \resizebox{.9\hsize}{!}{\includegraphics{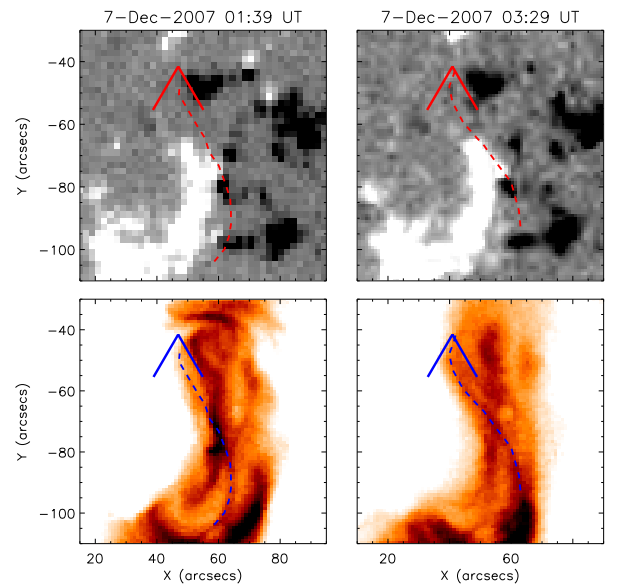}}
      \caption{\textsl{SOHO}/MDI {\it(top row)} and \textsl{Hinode}/XRT
               {\it(bottom row)} observations of AR~10977 
               showing the inversing crossing of sigmoidal threads on 7
               December at 01:39 and 03:29 UT. In the second panel the
               MDI data were taken at 
               7 December 03:12 UT and rotated to the time of the XRT image at
               03:29 UT. The MDI data are displayed between
               $\pm\!100$~Gauss. The little bipole seen on 7~December 
               after 03~UT at $(x,y)=(135,-130)$ and the corresponding
               set of compact X-ray loops (Figs.~\ref{fig:mdi} and
               \ref{fig:sigmoidform}) were used to coalign the images.
               An accuracy of approx. $1.5\arcsec$ is achieved by using
               this feature as well as other small scale features.}
         \label{fig:pilcrossing}
   \end{figure}

   \begin{figure*}
   \sidecaption
   \resizebox{.80\hsize}{!}{\includegraphics{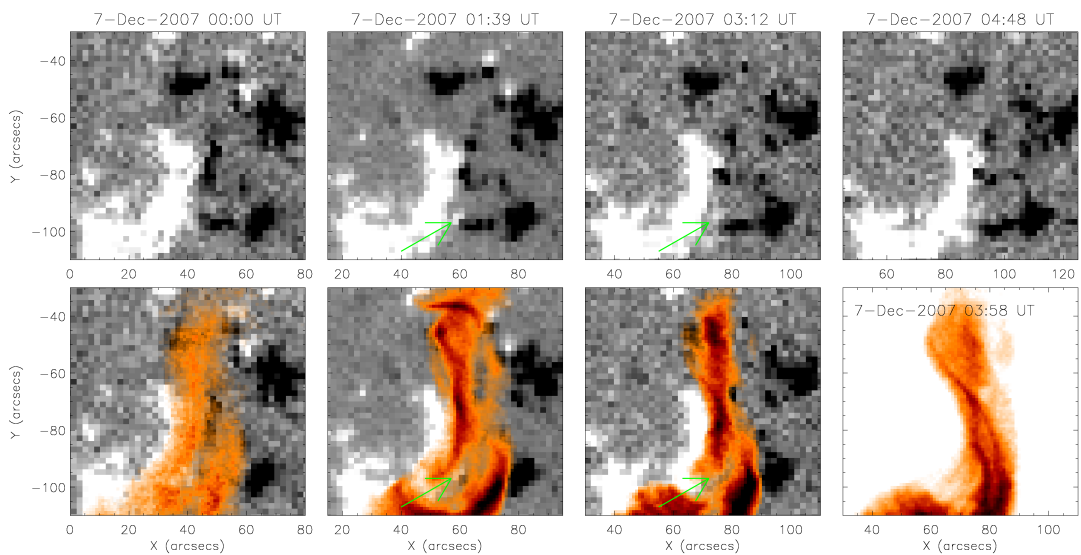}}
      \caption{XRT and MDI overlay {\it(bottom row)} showing that part of
      the linear feature merges with the continuous S shaped sigmoid as a
      consequence of the westward motion and partial cancellation of the
      flux at its southern end (marked by arrows). The fourth panel
      shows the sigmoid shortly before the eruption. There is no
      corresponding MDI image available. The same MDI images, including
      the subsequent one, taken during the eruption, are displayed in
      the top row for clarity.}
         \label{f:zoom}
   \end{figure*}

The evolution of the coronal magnetic field structure is studied using
data from the X-ray Telescope (XRT) onboard \textsl{Hinode} 
\citep{golub07} taken with the C~Poly filter, which image plasma 
between 2 and 10~MK. The active
region loops had an arcade structure during the emergence phase which
became increasingly sheared as the region evolved. During the decay
phase the soft X-ray loops became more aligned to the PIL in the
northern section of the region, whilst more potential-like arcade
field remained in the south (Fig.~\ref{fig:sigmoidform}). The major
flux cancellation episode took place in the northern region where the
shear was increasing.

A diffuse, but already continuous, forward S sigmoid was observed by 
6~December 2007 15:51~UT (Fig.~\ref{fig:sigmoidform}, panel~4).
The foward S shape indicates that the region had positive helicity
\citep{pevtsov97}. 50~minutes before the eruption, at 03:29~UT on
7~December, the sigmoid extends further to the north and is more defined
(Fig.~\ref{fig:sigmoidform}, panel~6). 
Figure~\ref{fig:pilcrossing} shows that the central part of
the sigmoid, while largely running along the PIL, crosses the PIL
in the inverse direction at $y\approx-75\arcsec$. At this location
the PIL bulges out to the west from its overall diagonal orientation
parallel to the middle section of the sigmoid. Much of the 
cancellation that leads to the
formation of the sigmoid is occurring in this area.

The inverse PIL crossing indicates that a flux rope has formed, but
is not conclusive evidence, since S shaped field lines with an inverse
PIL crossing in the middle can also form when an arcade is sheared
\citep{Antiochos&al1994}. However, since the sigmoid formed as a
consequence of shearing \emph{and} converging motions of similar
strength, which led to strong flux cancellation, the formation of a flux
rope structure is highly likely \citep{vanballegooijen89}. 
The configuration would be better described
as a (tangled) arcade only in the case that all new, initially
helical magnetic connections relax upwards into largely concave-down
coronal loops. Since this is unlikely, due to the accumulation of
axial flux above the PIL by the cancellation (Sect.~\ref{sss:flux}),
we will adopt the flux rope interpretation in the following. 
Further discussion of the flux
rope's magnetic structure is given in Sects.~\ref{ss:topology} and
\ref{sss:coherence}.

The sigmoidal region also has a linear bar-like feature aligned almost
north-south and crossing the sigmoid centre and the PIL in the normal
direction (panels~4 to 6 in Fig.~\ref{fig:sigmoidform}). The direction
of crossing of the PIL indicates that the linear feature is highly
sheared arcade-like field (i.e., not bald patch) passing \emph{over} the
forming flux rope in this phase of the evolution.

In the time leading up to the eruption of the sigmoid, 
there is a westward
motion of the southern end of the linear feature which is
situated at a latitude of roughly $-90\arcsec$ to $-100\arcsec$
and is composed of several, only marginally resolved threads.
Figure~\ref{f:zoom} shows that the footpoint motion is
occuring during a phase of westward motion and at 
the beginning of the dispersal
of a strong concentration
of positive flux in which the field lines are apparently rooted. These
motions appear to be driven by the development of a superganular cell
(see also panels~4 to 6 in Fig.~\ref{fig:mdi}). The positive flux
collides with a small negative polarity fragment which cancels during
the time period 7~December 00:00 to 06:24~UT. At 7~December 01:39~UT a
very small positive fragment has detached from the main concentration
at $-95\arcsec$ (marked by arrows in Fig.~\ref{f:zoom}).
It is cancelling with the negative polarity fragment
and is largely gone by 04:48~UT.
The southern end of a thread in the linear feature follows the path
of this positive flux fragment; see the panels at 03:12~UT in
Fig.~\ref{f:zoom}, at which time it has become the brightest thread in
the feature. The cancellation of the photospheric flux at the southern
end of this thread implies that the coronal flux in the thread must join
the sigmoid (and then be rooted somewhere along the section of the
sigmoid south of the cancellation site). Indeed, by the time the small
positive flux fragment has cancelled nearly completely, this thread has
formed a new connection with a prominent thread in the sigmoid (see the
XRT image at 03:58~UT in Fig.~\ref{f:zoom} and also the image 20~minutes
later in Fig.~\ref{fig:sigmoiderupt}).

Since the cancelling small positive flux patch is far smaller than the
negative flux patch at the northern end of the linear feature, only part
of the flux in the linear feature merges with the sigmoid in this phase.
The other, probably larger part must remain rooted in the westward
moving and dispersing major positive flux patch, which is also clear
from the continued presence of threads in the linear feature that end in
this area. However, the temporal association with the onset of the CME
at about 04:20~UT makes it quite likely that the partial transformation
of the arcade-like flux in the linear feature into helical flux in the
sigmoid represented the final step toward the destabilization of the
configuration.
The cancellation of the positive flux patch continues through and beyond the
eruption (see Figs.~\ref{fig:mdi} and \ref{f:zoom}).

The flux added to the sigmoid by the transformed part of the linear feature
must run under the magnetic axis of the forming flux rope at the position of
the cancellation ($y\approx-95\arcsec$) and above the axis where the linear
feature crosses the PIL in the normal direction, which is near the
position of the sigmoid's inverse PIL crossing ($y\approx-75\arcsec$).
Being wound about the forming flux rope's axis, at least this part of
the linear feature must follow the motion of the axis in the subsequent
eruption.

\begin{figure}
   \centering
   \resizebox{\hsize}{!}{\includegraphics{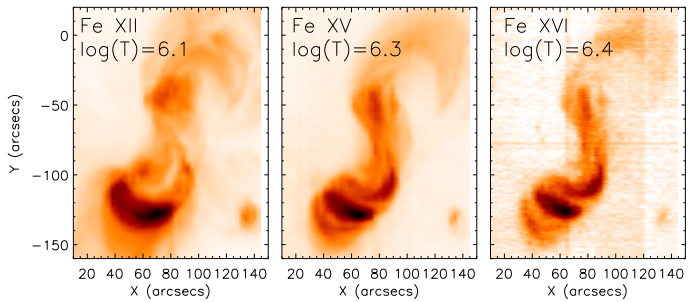}}
      \caption{\textsl{Hinode}/EIS observations of AR~10977 showing the
               temperature structure of the sigmoid and linear
               feature. The EIS scan is built from west to east and
               was taken between 7~December 03:27~UT and 04:19~UT with
               the 1\arcsec\ slit in step sizes of 3\arcsec.}
         \label{fig:eis}
   \end{figure}

The temperature structure of the region is studied using the Extreme
ultra-violet Imaging Spectrometer (EIS) onboard \textsl{Hinode}
\citep{culhane07}. Figure~\ref{fig:eis} shows simultaneous images
in three lines with different formation temperature built from an EIS
raster scan immediately before the onset of the CME. 
The sigmoid is seen at temperatures
$\log(T[\mathrm{K}])=6.1$ to 6.4, but the linear feature is only
seen in the hotter lines imaging plasma at $\log(T[\mathrm{K}])=6.3$
to $6.4$. The EIS data do include flare lines
which image at higher temperatures, but the emission is weak at these
wavelengths and does not allow the study of the linear feature versus
the sigmoid. The high plasma temperature supports the view that part of
the linear feature joins the sigmoid through the reconnection that
enables the cancellation of the positive flux at its southern end.
A sigmoid study by \citet{Tripathi&al2009} showed that plasma 
near the axis of the inferred
flux rope was cooler than the double-J shaped sigmoid at its edge.
This suggests that the linear feature in the present event does not 
follow a passive bundle of flux near the magnetic axis of the 
flux rope at this stage of the evolution.

\begin{figure}
   \centering
   \resizebox{.85\hsize}{!}{\includegraphics{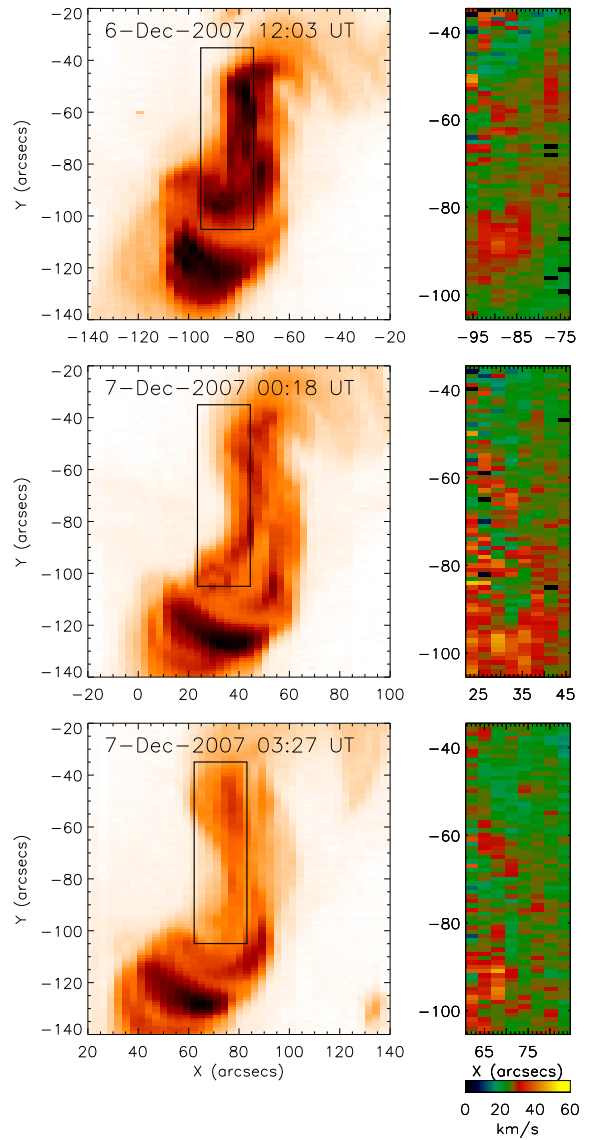}}
      \caption{\textsl{Hinode}/EIS observations of AR~10977 in the
               Fe~XV 284~{\AA} line formed at temperature 2~MK.
	       The three EIS scans were taken with the same observation
	       parameters; their start times are given.
	       The left panels display Fe~XV radiance and the right
	       panels display the non-thermal component of the
	       line width in the sub-area indicated.}
         \label{f:linewidths}
   \end{figure}

Broadenings of the Fe~XV line, which show both the sigmoid and the
linear feature clearly, support this view. Figure~\ref{f:linewidths}
presents maps of Fe~XV radiance and the non-thermal component of
the line width (FWHM) from the three
available EIS raster scans prior to the eruption which included this
line. The non-thermal velocities are calculated using the IDL 
Solar Software
routine \texttt{eis$\_$width2velocity} which removes instrumental width,
taken to be 0.057~\AA~for the $1\arcsec$ slit \cite{brown08},
and thermal Doppler width.
The broadenings trace out the diagonal part of the S shaped sigmoid and
the linear feature (the whole feature in the first two maps and mainly
its southern end in the third map). The implied microturbulence
indicates strongly that both structures are sites of ongoing energy
release.
For the S shaped sigmoid, this supports the now generally held view that
such structures form by current dissipation in separatrices
\cite[e.g.,][]{titov99}. For the linear structure, the line width
enhancements indicate that this flux interacts with surrounding flux or
experiences internal changes. Both are likely, due to the westward
motion and due to the dispersal and partial cancellation of the positive
flux at the feature's southern end. Accordingly, the line is broadest in
this location in the two maps on 7~December, particularly in the map at
03:27~UT. Enhanced widths correlate largely, but not fully, with
enhanced radiance of this relatively ``hot'' line. The third map shows
the largest widths in an area of weak emission, and the bright arcade
south of the sigmoid (outside the line width maps) has line widths close
to the background level within the field of view of the maps. 
Considering that the EIS scan takes 52 minutes to build the image 
of the active region, the location of the cancellation
site at $y\approx-95\arcsec$ will have moved from $x\approx75\arcsec$ 
(as seen in the MDI image at 03:12 UT) to $x\approx81\arcsec$ 
as seen in the EIS image which reaches the location of cancellation
at  03:47 UT. 
The cancellation site shows line width enhancements similar to
the main part of the linear feature and the continuous S shaped sigmoid
thread, but it does not stand out as a particular enhancement. This
may have several origins: most of this cancellation event proceeds
between the two EIS scans; it involves only relatively a small amount of
flux; the associated reconnection occurs low in the atmosphere so that
the coronal responses remain weak.

\section{Eruption}
\label{s:eruption}

  \begin{figure}
   \centering
   \resizebox{\hsize}{!}{\includegraphics{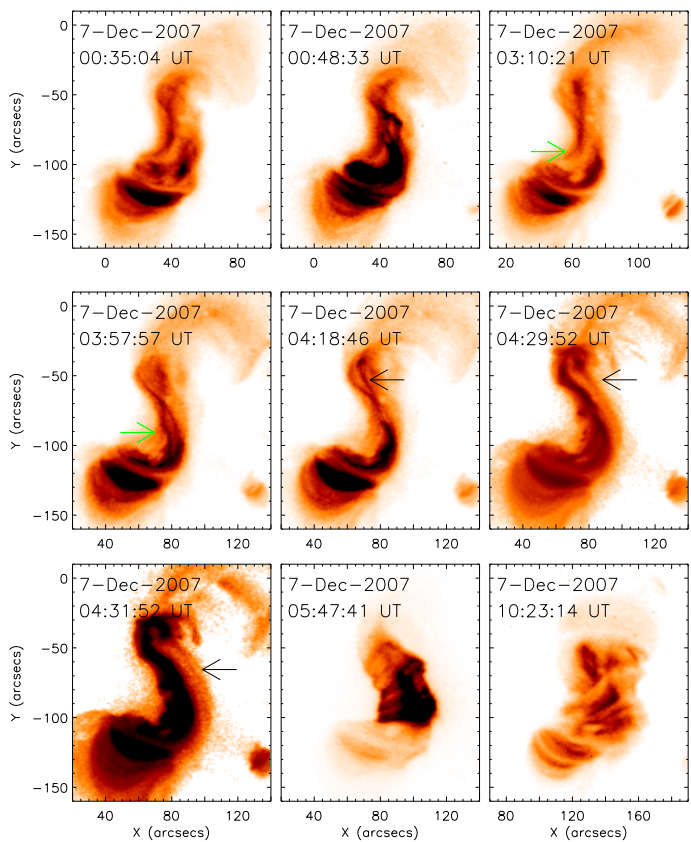}}
      \caption{\textsl{Hinode}/XRT C Poly filter observations showing
               the failed eruption {\it(top row)} and the successful eruption
               {\it(second and third rows)}, including the post-eruption
	       configuration (final two panels). Green arrows in panels~3 and 4
               indicate the motion of the southern footpoint of the
               linear feature. Black arrows in panels~5 to 7 indicate
               the eruption of the linear feature. Panels 6 and 7 have
               been log displayed to bring out this faint feature.}
         \label{fig:sigmoiderupt}
   \end{figure}

The eruption of the flux rope can be seen to occur in two steps; a
failed eruption followed 3.5 hours later by a successful eruption
which produces the white light CME.

\begin{figure}
   \centering
   \resizebox{\hsize}{!}{\includegraphics{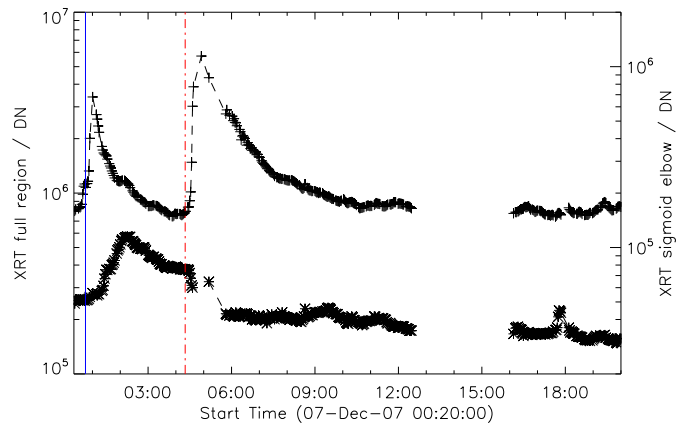}}
      \caption{Light curves of the soft X-ray emission from the whole
               active region (crosses, top plot) and the northern
               elbow only (asterisks, bottom plot).
               The failed eruption beginning around 7~December
               00:45~UT (blue solid line) produces a temporary
               enhancement in active region emission and is followed
               by an enhancement in the northern arm of the sigmoid.
               The main eruption (red dash-dot line) is accompanied by
               enhanced emission in the centre of the active region
               beginning at 7~December 04:20~UT and reduced emission
               in the northern elbow.}
         \label{fig:lightcurves}
   \end{figure}

The failed eruption occurs while the linear feature 
begins to transform into
the sigmoid and may facilitate the reconnection between the two
flux systems. It commences around 7~December 00:45~UT and is seen as the
rise of a bundle of field lines at the sigmoid
centre (Fig.~\ref{fig:sigmoiderupt}, panel~2 at 00:48~UT). After the
failed eruption the active region's overall structure remains the
same, but the northern sigmoid elbow (where the top of the S curves
around) has developed and has increased soft X-ray emission
(Fig.~\ref{fig:lightcurves}). The XRT images during 03:57--04:19~UT in
Fig.~\ref{fig:sigmoiderupt} show two superimposed sigmoids; 
one which has the large northern elbow and a second, brighter S shaped
emission trace, which branches off at latitude $\approx-40\arcsec$.

The successful (ejective) eruption begins in the decay phase of the
failed eruption, at approximately 04:20~UT on 7~December 2007
(Fig.~\ref{fig:lightcurves}). It
is accompanied by a B1.4-class \textsl{GOES} flare and an EUV wave
\citep{Ma&al2009}. 
The CME, observed by \textsl{SOHO}/LASCO,
displayed a flux rope structure and propagated westward with a projected
velocity of $\approx\!300$~km\,s$^{-1}$. The true velocity may have been
much higher, since the ejection originated near sun centre.

In the lower corona, the
eruption is evidenced by the motion of the linear feature as seen in
soft X-rays. This feature, although it is now increasingly curved,
can be identified by the isolated negative flux patch at its 
northern end and moves to the west as does the associated 
CME (see black arrows in panels
5--7 of Fig.~\ref{fig:sigmoiderupt}). Thus, as argued in
Sect.~\ref{sect:sigmoidformation}, the feature now moves with
the rising flux rope's axis. This is essentially in line with
the earlier interpretation of bar-like features in \citet{moore01} and
\citet{mckenzie08}.

Comparing the properties of the linear feature with the current
layer in arcade field overlying an erupting flux rope in the simulation
by \citet{aulanier10}, we find differences as well as common aspects.
The observed linear feature was present already before the eruption,
while the current layer in the simulation formed as a result of the
eruption. It rotated in the counterclockwise direction,
opposite to the rotation in the simulation for the same chirality of the
active region field. Similar to the simulation, the observed linear
feature represents a narrow volume of enhanced currents, as indicated by
its high temperature and enhanced Fe~XV line width. However, we find it
to be associated with the overlying arcade-like field only in the phase
prior to the eruption, while it must be a part of the flux rope in the
course of the eruption. Some of these differences are a consequence of
the different formation timescales of the flux rope. In the simulation,
the flux rope is fully developed prior to the eruption and the currents
within it decrease while it expands in the eruption. In the event
studied here, the overlying flux that threads the linear feature joins
the flux of the growing rope only immediately prior to and in the course
of the eruption. Thus the linear feature can be related to enhanced
currents in overlying field prior to the eruption and to enhanced
currents in the flux rope in the course of the eruption.

Under the erupting structure the first, highly sheared post-eruption
(flare) loops appear. All of these run very closely along the
northern section of the sigmoid, passing over the location of the
inverse PIL crossing where the major flux rope formation episode has
taken place. They branch off from the central S trace of the sigmoid in
the range $y\approx-100\arcsec$ to $-80\arcsec$. In contrast, the loops
of the flare arcade formed in the gradual phase of the event cross the
PIL at nearly right angles (see panels 6 to 8 in
Fig.~\ref{fig:sigmoiderupt}).

These data indicate clearly that flare reconnection completed the
transformation of the flux in the linear structure into the erupting
flux (which is to be expected because the linear feature passed over the
centre of the erupting structure immediately before the event). As a
result of such reconnection, flare loops must end in the positive flux
concentration where the linear feature was rooted before the eruption.
Exactly this is the case, starting with the first flare loops at
04:30--32~UT early in the impulsive phase (Fig.~\ref{fig:sigmoiderupt}).
Explosive chromospheric evaporation, a further sign of reconnection,
commences in this area at $(x,y)\approx(70\arcsec,-85\arcsec)$ between
04:30 and 04:35~UT, i.e., also early in the impulsive phase
\citep{Chen&Ding2010}.

A complementary view is obtained by asking where the flux in the
erupting rope was rooted. The flare brightenings are largely associated
with the smaller and brighter of the two sigmoids, whose northern elbow
encloses the strong patch of negative flux at $y\approx-40\arcsec$ to
$-50\arcsec$, where the linear feature is rooted at its northern end.
Figures ~\ref{fig:mdi} and \ref{f:zoom} show that this is a largely
isolated flux concentration. The flux of the erupting rope must largely
be rooted in this patch. Hence, the erupting, and now curved, 
linear feature has become part of the erupting flux rope. 

\begin{figure}
   \centering
   \resizebox{\hsize}{!}{\includegraphics{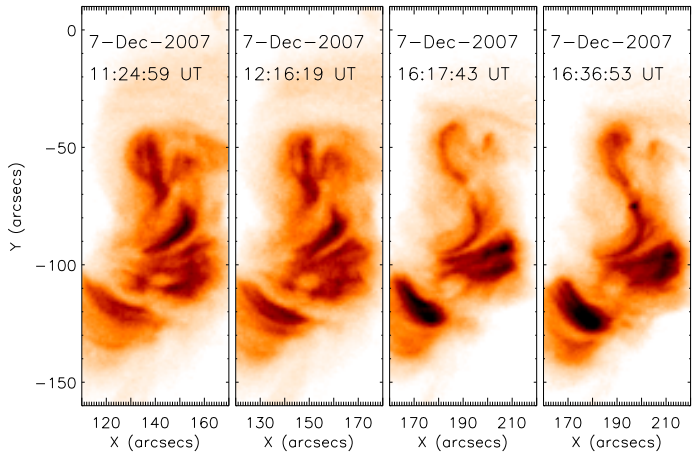}}
      \caption{\textsl{Hinode}/XRT C Poly filter observations showing the
               reformation of the sigmoid 
               after the successful eruption which produced a CME.}
         \label{f:reformation}
   \end{figure}

It appears that a trace of the sigmoid remains visible even under the
arcade. However, since the region is near Sun centre, this feature
could also result from enhanced brightness at the top of the flare loops,
which occurs in some events. 
The X-ray images immediately after the eruption do not show any clear
indications of a surviving sigmoid. Rather, the sigmoid was replaced by
two sets of loops, the usual nearly potential arcade in the southern
half and a set of highly sheared loops in the northern half
(Figs.~\ref{fig:sigmoiderupt} and ~\ref{f:reformation}). 
The presence of substantial
shear immediately after the eruption could indicate that part of the
axial flux in the rope survived in the northern section, but it could also
result from the high shear of the reclosing field which is indicated by
the shape of the first flare loops. The data do not appear to provide a
clue which of the two options applies here.

The dimmings following the eruption are not well defined and do not
allow a reliable determination of their flux content. 

\section{Sigmoid reformation}
\label{s:reformation}

One of the dominant loops in the northern area of the destroyed
sigmoid begins to develop a half S (or J) shape very soon after the
flare arcade has faded; see the image at 11:25~UT in
Fig.~\ref{f:reformation}. The following hour leading up to the XRT data gap
between 12:28--16:02~UT sees a gradual approach of the southern end of this
structure with the J shaped nearest loop in the southern part of
the post-event arcade. The images after the data gap show that
the two loops merge into a new, continuous S shaped sigmoid in the same
location as the pre-eruption sigmoid, albeit smaller 
(Fig.~\ref{f:reformation}, panels~3 and 4). The evolution
continues to be dynamic with
the new sigmoid showing reduced brightness in the middle and a bright
spot near this intensity dip for part of the time, as well as a changing
intensity ratio between the northern and southern arms and changing
sharpness, with a multi-thread appearance at times.

The reformation of the sigmoid and its subsequent dynamics are driven by
the changes in the photospheric flux distribution. In particular, the
merging into a new continuous S shaped sigmoid is cospatial with the
ongoing flux cancellation episode in the range $y\approx-65\arcsec$ 
to $-85\arcsec$
(see the final MDI frame in Fig.~\ref{f:zoom}). Although the amount of
cancelled flux in the approximately 6 hours of sigmoid 
reformation is far smaller
than the flux cancelled in the 2.5~days prior to the eruption
(Figs.~\ref{fig:mdigraph} and \ref{fig:mdi}), a complete new sigmoid is
formed. We will discuss this difference in Sect.~\ref{sss:flux}.

The new sigmoid was destroyed in a very weak eruption, including a CME,
on 8~December after 17~UT.


\section{Discussion}
\label{s:discussion}


\subsection{Flux rope topology}
\label{ss:topology}

The observations of AR~10977 strongly support the picture of the
gradual transformation of arcade field into a flux rope as a result of
photospheric flux cancellation over extended periods prior to an
eruption, as suggested by \citet{vanballegooijen89}. A major \emph{flux
cancellation episode} occurs in the period of 2.5~days between the time
of peak flux content after the region's emergence and a CME/flare
event. It comprises many \emph{cancellation events} of flux fragments at
the internal PIL. This transforms the initially
weakly sheared arcade of soft X-ray loops into a sigmoidal region
exhibiting a key signature of a flux rope formed by cancellation:
the inverse crossing of the PIL by the middle of the sigmoid,
bracketed by two regular PIL crossings by the sigmoid elbows.

The specific magnetic structure of a flux rope can be split into two
categories; flux ropes having their underside rooted in the dense lower
atmosphere having a bald  patch separatric surface (BPSS) topology
\citep{titov99}, and flux ropes which are situated in the corona  having
a so-called hyperbolic flux tube (HFT), i.e., a magnetic X-type
structure, at their underside \citep{Titov&al2002}.

If the reconnection that enables flux cancellation proceeds in or very
near the photospheric level, as suggested by \citet{vanballegooijen89},
then it is likely that the flux rope initially forms with BPSS topology.
However, in the course of a cancellation episode, the topology can
change into one containing an HFT \citep{aulanier10}. The transition
begins by splitting the bald patch into two sections, with the HFT
connecting their inner end points through the corona. The two bald patch
sections shrink while the HFT grows in the course of the transition.

BPSS and HFT flux ropes exhibit a completely different behaviour in the
event of an eruption. A BPSS flux rope must split in two along a
horizontal line, with only the upper part erupting, since its lower part
is tied to the lower solar atmosphere \citep{gibson06}. Since sigmoids
form in the lower surface of their host flux ropes \citep{titov99,
green07}, a sigmoid in a BPSS flux rope forms in the bottom part of the
BPSS and is likely
to survive the eruption, at least partly. These sigmoids are located
below the developing flare current sheet, hence also below the
post-eruption arcade. Sigmoids forming in the quasi-separatrix layer in
the underside of an HFT flux rope are located in and above the HFT. In
an eruption, the HFT pinches into the flare current sheet
\citep{Torok&al2004}, so that essentially the whole flux rope above it
is ejected and the sigmoid ceases to exist after the eruption.

From the destruction of the sigmoid in the eruption on 7~December 2007
we can thus conclude that the flux rope did likely possess an HFT
structure at this time.


\subsection{Magnetic flux budget}
\label{ss:fluxbudget}

The unsigned flux content of the active region at the time of the
eruption is $\approx\!1.17\times10^{21}$~Mx. Up to this point,
a total flux of $0.71\times10^{21}$~Mx is cancelled at the internal PIL,
equal to $\approx\!34\%$ of the peak negative flux in the active region
prior to the cancellation episode. If all flux lost
at the internal PIL caused other flux of the active region
to transform into the flux rope in the manner anticipated by
\citet{vanballegooijen89}, then $\approx\!60\%$ of the active region
flux is in the rope at the time of the eruption. The ratio would be
even higher if the flux crossing the PIL south of the sigmoid were
excluded from the unsigned flux content of the active region. 
A flux ratio of this
magnitude can still permit force-free equilibrium in model fields (see
Fig.~1 in \citeauthor{Valori&al2010} \citeyear{Valori&al2010} for
examples). However, it differs strongly from the threshold value for
the occurrence of force imbalance found in recent numerical
investigations of flux rope structure and stability in decaying active
regions.

By inserting a flux rope above and along the PIL in the extrapolated
potential field of active regions that have undergone significant flux
cancellation, and through subsequent numerical relaxation of the
configuration, \citet{bobra08}, \citet{Su&al2009} and
\citet{savcheva09} found a limit for the axial flux in a rope that can
be held in stable force-free equilibrium by the other flux of the
active region. The limit amounts to at most 10\% of the unsigned flux
in the active region for the three regions studied in \citet{bobra08}
and \citet{Su&al2009}. The limiting axial flux given in
\citet{savcheva09} for a fourth, strongly decayed and sigmoidal active
region combines with the amount of unsigned flux in the region (which
we find to be $3.5\times10^{21}$~Mx) to a limiting ratio of
$\approx\!14\%$.

These three investigations also indicate that the poloidal flux in the
rope is not well constrained by their method of comparing field lines
in an array of relaxed flux rope configurations with varying axial and
poloidal fluxes to observed coronal features. Typically, the poloidal
flux could be raised from the best-fit value in a wide range of about
an order of magnitude, whilst still fitting the observed structures, 
until the numerical relaxation failed. Expressing the average twist
angle in the flux rope as
$\Theta=\pi\Phi_\mathrm{pol}/\Phi_\mathrm{axi}$, the rope is expected
to be stable against the helical kink mode for a poloidal flux of
(typically) up
to $\Phi_\mathrm{pol}\sim3\Phi_\mathrm{axi}$ \citep{Torok&al2004,
Fan&Gibson2004}. Indeed, \citet{bobra08} find a stable configuration
with $\Phi_\mathrm{pol}/\Phi_\mathrm{axi}=2.6$ (their model~7 for
AR~9997/10000) and \citet{savcheva09} even find a stable configuration
with $\Phi_\mathrm{pol}/\Phi_\mathrm{axi}=4.8$ (their model for a
sigmoid on 12~Feb 2007, 08:38~UT). These values indicate that only
$\sim\!1/4$ of the flux in the rope may be axially directed,
potentially reducing the discrepancy with our result to a large
extent. However, the flux ropes that best fitted the observed active
region structure in these investigations were systematically found to
have considerably lower amounts of poloidal flux,
$\Phi_\mathrm{pol}/\Phi_\mathrm{axi}\ll1$ in \citet{bobra08} and
\citet{Su&al2009} and $\sim\!1/4$ in \citet{savcheva09}, so that other
explanations for the discrepancy should be considered.

Observations of interplanetary CMEs which exhibit a flux rope
structure indicate a dominance of the poloidal flux over the axial
flux, typically by a factor $\sim\!3$ \cite[e.g.,][]{mandrini05,
attrill06, qiu07}. However, much of this flux is added to the rope by
reconnection after the launch of the CME \citep{qiu07}. Therefore,
these observations do not provide conclusive evidence of the partition
between axial and poloidal flux in the rope at the onset of the
eruption.

We note that \citet{Sterling&al2010} recently also found a
considerable amount of cancellation prior to an eruption in another
active region. About 20\% of the unsigned active region flux cancelled
over two days preceding the event, and further cancellation may have
occurred at earlier times not analysed in that study.

In the following, we discuss two aspects of the process of flux rope
formation by cancellation which may help reconcile the discrepancy
between our measurements of the amount of flux at the onset of
eruption  in Sect.~\ref{sect:mdi} and the results of the numerical
modelling based on the flux rope insertion method.


\subsubsection{Cancelled flux versus rope flux}
\label{sss:flux}

   \begin{figure*}[t]
   \centering
   \sidecaption
   \resizebox{0.6\hsize}{!}{\includegraphics{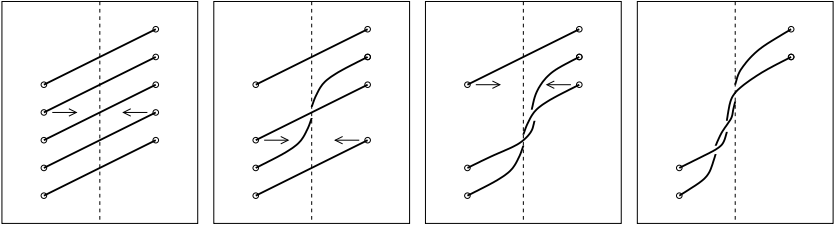}}
      \caption{Schematic of a cancellation episode which consists of three
      cancellation events and transforms the part of the initial arcade
      shown in the first panel into a flux rope. The PIL is drawn dashed.
      The ratio of footpoint displacement of the sheared arcade loops to
      the length of the active section of the PIL (where cancellation
      proceeds) is $2:3$. Correspondingly, $2/5$ of the flux
      is transformed into the rope, while $3/5$ cancel.}
         \label{f:schematic1}
   \end{figure*}

\citet{vanballegooijen89} argued that magnetic flux can submerge below
the photosphere only if it forms small loops with a radius of
curvature comparable to the photospheric pressure scale height, giving
an estimated maximum distance between the loop footpoints of
$\sim\!900$~km. If the flux in the active region has shear, most loops
in the original arcade will have a footpoint separation in the
direction of the PIL larger than this value. This is definitely the
case for AR~10977  from the onset of the cancellation episode (see
Fig.~\ref{fig:sigmoidform}, panel~1). When the footpoints of these
loops are transported toward the PIL, the loops remain too long for
submergence. Rather they reconnect with other loops at, or somewhat
above, the photospheric PIL, creating the short loops which can submerge and
long loops rooted away from the cancellation site at the remote footpoints
of the reconnecting loops. The downward tension force of the long
loops is insufficient to drive them below the photosphere, and their
middle part tends to be more aligned with the PIL than the original
two loops. Therefore, cancellation leads to an accumulation of flux
which runs along the \emph{active section} of the PIL, where flows
converging toward the PIL drive the cancellation episode and where
sufficient flux exists on both sides to enable cancellation events.

If axial flux  already exists above the PIL (which can result from
previous cancellation) then the newly reconnected long loops are
helical. They wrap around the bottom side of the existing axial flux
such that a flux rope gradually forms as the cancellation episode
continues. 

If only a single cancellation event occurs, then, by the nature of the
process, the flux accumulated at the PIL equals the cancelled flux.
However, this is no longer the case when subsequent cancellation
occurs which involves the product of a previous event. In this case,
no further flux is added, rather the accumulated flux will extend
further along the PIL. An amount of flux equal to the flux in the long
loop is lost from the system. This is illustrated by the schematic in
Fig.~\ref{f:schematic1}. The effect can also be seen in the filament
formation scenario proposed in \citeauthor{Martens&Zwaan2001}
(\citeyear{Martens&Zwaan2001}, their Fig.~6), but has not been
explicitly addressed in that paper.

Thus, of the original loops that pass over the active section of the
PIL, three categories must be distinguished: loops rooted at one end
in flux that cancels, loops rooted at both ends in flux that cancels,
and loops rooted in non-cancelling flux at sufficient distance to the
PIL. Half of the flux in the first category will be transformed into
the flux that accumulates at the PIL and eventually forms a flux rope,
while the other half cancels. The flux in the second category will be
lost without contributing to the amount of flux in the rope. The third
category remains in the active region as overlying, stabilizing flux.

As can be seen from Fig.~\ref{f:schematic1}, the ratio between  the
flux accumulated at the PIL and the cancelled flux depends on the
shear of the original arcade and on the length of the active section
of the PIL. For roughly uniform flux along the active
section of the PIL, it is given by the ratio between the
displacement of the loop footpoints parallel to the PIL and the length
of the active section (with an upper limit of
unity), remaining small for a weakly sheared arcade.

This ratio can change considerably in the course of a cancellation
episode, by changing shear as well as by changing length of the active
PIL section. Typically, the shear increases and the length of the active
PIL section decreases when large parts of the flux in an active region
cancel. Thus, the process of flux transformation from arcade to flux
rope due to cancellation accelerates. Therefore, it seems plausible that
sigmoids are typically observed late in a cancellation episode, closer to
an eruption than to the onset of the cancellation, as in the event studied
here, and that the occurrence of sigmoids has such a high correlation
with eruptive activity \citep{canfield99}.

Figure~\ref{fig:mdi} shows that the shear increases while the flux in
AR~10977 cancels. The change in the length of the active PIL section is
less pronounced. The shear remains weak during 4~December, but grows
considerably on 5~December, which is also clearly visible in the soft
X-ray images in Fig.~\ref{fig:sigmoidform}. Still, the shear remains
moderate until about mid-day on 5~December, by which time
$\approx\!40\%$ of the cancellation has occurred
(Table~\ref{table:1}). Most of this flux and part of the subsequently
cancelled flux is lost from the active region without contributing to
the amount of flux in the rope. This reduces the discrepancy to the
numerically obtained stability limit by a factor $\sim\!2$,
or somewhat higher.

The sigmoid reformation after the CME started in a region of highly
sheared flux, as evidenced by the northern set of post-eruption loops
(images at 10:23--11:25~UT in Figs.~\ref{fig:sigmoiderupt} and
\ref{f:reformation}). Therefore, a new flux rope and sigmoid could be
formed through an amount of cancellation much less than the cancellation
in the period prior to the eruptions on 7~December.


\subsubsection{Coherence of the flux rope}
\label{sss:coherence}

A further cause for the discrepancy may be found in the differing
degree of coherence in the flux rope structure. The flux rope inserted
in the three numerical studies mentioned above is of course fully
coherent along its length and unlikely to lose much of this coherence
in the relaxation. (Since a vacuum sheath surrounding the flux rope
was also inserted, both the external incoherent field and the inserted
flux rope could expand in the relaxation to form a current layer,
where much of the structural differences between the flux rope and its
exterior could be resolved, so that these did not strongly propagate
into the flux rope.) In order to form such a coherent flux rope by
cancellation, the cancellation must proceed largely uniformly along a
substantial fraction of the length of the resulting rope. Only then
will concave-up field lines exist along a substantial fraction of its
length,  and only then will field lines spiral \emph{above} the middle
part of the rope (these field lines must be concave-up near one or
both ends of the rope). The observations indicate that the
cancellation episode evolved from an initially more uniform
distribution of cancellation events along the active section
of the PIL to a more localised pattern (compare panels 1--3 in
Fig.~\ref{fig:mdi} with panels 4--5). At the same time, the length of
the sigmoid increased. The cancellation was by far strongest in the
middle of the forming sigmoid where the inverse PIL crossing occurred.
At this position the number of helical field lines running under the
axis of the forming flux rope is \emph{much} larger than the number of
helical field lines passing over the axis.

As discussed in Sect.~\ref{sect:sigmoidformation}, the linear bar-like
feature is suggestive to be of arcade structure prior to the eruption,
simply passing over the gradually forming flux rope without already
being a part of it. It passes over the sigmoid's inverse PIL crossing,
where most rope field lines are upward concave. By partly joining the
sigmoid, the linear feature adds downward
concave flux in this location, leading to a more complete and more
coherent flux rope structure. The temporal association of this process
with the onset of the successful eruption is likely no coincidence.
Also, the coherence of the flux rope is likely completed only in the
process of the eruption.


\section{Summary and conclusions}
\label{s:conclusions}

We present observations of NOAA active region 10977 which was 
observed from emergence through to the decay phase, during which a
coronal mass ejection occurred. The evolution of the magnetic field in
the decay phase was dominated by  fragmentation of the photospheric
field, the development of supergranular flows, and cancellation of
photospheric flux fragments at the developing supergranular
boundaries. The observations support the interpretation that a major
flux cancellation episode was responsible for the formation of a flux
rope in the active region during the $\sim\!2.5$~days preceding the
CME. A soft X-ray sigmoid, crossing the PIL in the inverse direction,
suggests that a flux rope
has formed \emph{prior to} the eruption. The sigmoid was
destroyed by the eruption, indicating that the flux rope had developed a
hyperbolic flux tube (HFT) at its underside. Continued cancellation
through and after the eruption led to the quick reformation of the
sigmoid, which later erupted again.

The unsigned flux in the active region at the time of the eruption is
about $1.17\times10^{21}$~Mx. During the 2.5~day period before the 
eruption about $0.71\times10^{21}$~Mx cancels at the internal PIL  of
the active region, equal to $\approx\!34\%$ of the peak negative flux
in the active region prior to the cancellation episode.
If the reconnection which enables the flux
cancellation transformed an equal amount of arcade flux into the flux
of the forming rope according to the mechanism put forward by
\citet{vanballegooijen89}, then the rope would contain about 60\% of
the active region flux. This is much larger, by a factor $\approx$\,4--6,
than the limiting axial flux value which would maintain force balance in
a decaying active region, as found in recent numerical modelling
\citep{bobra08, Su&al2009, savcheva09, aulanier10, Amari&al2010}.

We point out that the amount of flux transformed from the initial
arcade into the flux rope can differ considerably from the amount of
cancelled flux. If the initial arcade is only weakly sheared, then the
amount of arcade flux rooted beyond the ends of the active section of
the PIL is far smaller than the amount of flux at both sides of the
active section. The former is the flux available for transformation
into the flux rope. The latter can cancel without contributing to the
flux in the rope. Their ratio is (very roughly) given by the ratio
between the displacement of the arcade loop footpoints parallel to the
PIL and the length of the active section of the PIL,
limited to unity. Thus, for an initially only weakly sheared arcade,
the amount of flux cancelled can be much larger than the amount of flux
transformed into the flux rope.

If the cancellation occurs predominantly along a section of the  PIL
which is small in relation to the extent of the forming flux rope, as
in the late phase of the studied event, then most of the helical,
concave-up field lines have their bottom section in this localized
area. Helical field lines passing over this area are strongly
underrepresented, so that the resulting flux rope is incoherent to a
considerable degree. This is likely to contribute to the discrepancy
with the numerically obtained stability limit. Also, it suggests that
flux ropes built by flux cancellation often reach coherence only in
the course of an eruption.

The sigmoid forming in AR~10977 includes a linear, bar-like feature,
which is hotter than the S shaped threads prior to the eruption. In
this phase, the feature must be interpreted as a bundle of arcade field
lines passing over the developing flux rope. The data yield evidence
that this flux becomes part of the flux rope immediately before and
in the early stages of the eruption,
through the cancellation at its southern end and the further
reconnection which establishes the coherence of the flux rope.
Subsequently it traces the path of the erupting flux rope body, as
suggested for similar features in earlier events \citep{moore01,
mckenzie08}.

\begin{acknowledgements}
We thank the referee, G.~Aulanier, for the very constructive reports
that significantly helped improve the clarity of this paper and Duncan
Mackay for helpful discussions and comments on the manuscript.
LMG was supported by a Royal Society Dorothy Hodgkin Fellowship and a
Leverhulme Fellowship. 
AJW was supported by STFC via a PhD studentship. 
BK acknowledges support by the DFG, the STFC, and NASA grants
NNH06AD58I and NNX08AG44G.
\emph{Hinode} is a Japanese mission developed and launched by 
ISAS/JAXA, collaborating with NAOJ as a domestic partner, 
NASA and STFC (UK) as international partners.  Scientific 
operation of the \emph{Hinode} mission is conducted by the 
\emph{Hinode} science team organised at ISAS/JAXA.  This 
team mainly consists of scientists from institutes in the 
partner countries.  Support for the post-launch operation is provided by 
JAXA and NAOJ (Japan), STFC (UK), NASA (USA), ESA, and NSC (Norway).
This research has made use of the LASCO CME catalog, generated and
maintained at the CDAW Data Center by NASA and The Catholic University
of America in cooperation with the NRL. \textsl{SOHO} is a project of
international cooperation between ESA and NASA.
\end{acknowledgements}

\bibliographystyle{aa} 
\bibliography{references} 

\end{document}